\documentclass[twocolumn,preprintnumbers,amsmath,aps]{revtex4-1}
\usepackage{graphicx}
\usepackage{dcolumn}
\usepackage{bm}
\usepackage{subfigure}
\usepackage{color}
\usepackage{amssymb}
\begin{document}
%%%%%%%%%%%%%%%%%%%%%%%%%%%%
\newcommand{\kvec}{\mbox{{\scriptsize {\bf k}}}}
%%%%%%%%%%%%%%%%%%%%%%%%%%%%
\def\eq#1{(\ref{#1})}
\def\fig#1{figure\hspace{1mm}\ref{#1}}
\def\tab#1{table\hspace{1mm}\ref{#1}}
%%%%%%%%%%%%%%%%%%%%%%%%%%%%
\title{Electron-phonon coupling in Kekul{\' e}-ordered graphene}
%%%%%%%%%%%%%%%%%%%%%%%%%%%%
\author{Dominik Szcz{\c{e}}{\'s}niak}\email{d.szczesniak@ujd.edu.pl}
%%%%%%%%%%%%
\affiliation{Institute of Physics, Faculty of Science and Technology, Jan D{\l}ugosz University in Cz{\c{e}}stochowa, 13/15 Armii Krajowej Ave., 42200 Cz{\c{e}}stochowa, Poland}
%%%%%%%%%%%%
\date{\today} 
\begin{abstract}
%%%%%%%%%%%%%%%%%%%%%%%%%%%%%%%%%%%%%%%%%%%%%%%%%%%

Breaking the intrinsic chirality of quasiparticles in graphene enables the emergence of new and intriguing phases. One such paradigmatic example is the bond density wave, which leads to a Kekul{\' e}-ordered structure and underpins exotic electronic states where electron-phonon interactions can play a fundamental role. Here, it is shown that the relevant physics of these correlations can be resolved locally, according to the behavior of interatomic characteristics. For this purpose a robust distance-dependent framework for describing electronic structure of graphene with Kekul{\' e} bond order is presented. Given this insight, the strength of electron-phonon interactions is found to scale linearly with the electronic coupling, contributing to a uniform picture of this relationship in distorted graphene structures. Moreover, it is shown that the introduced distortion yields a strongly non-uniform spatial distribution of the pairing strength that eventually leads to the induction of periodically distributed domains of enhanced electron-phonon coupling. These findings help elucidate certain peculiar aspects of phonon-mediated phenomena in graphene, particularly the associated superconducting phase, and offer potential pathways for their further engineering.

%%%%%%%%%%%%%%%%%%%%%%%%%%%%%%%%%%%%%%%%%%%%%%%%%%%
\end{abstract}
\maketitle
%

%%%%%%%%%%%%%%%%%%%%%%%%%%%%%%%%%%%%%%%%%%%%%%%%%%%%%%%%%%%%%%%
\section{Introduction}
%%%%%%%%%%%%%%%%%%%%%%%%%%%%%%%%%%%%%%%%%%%%%%%%%%%%%%%%%%%%%%%

Graphene, the two-dimensional form of graphite, is renowned for numerous exceptional properties \cite{castro2009, bonaccorso2010, lee2008, balandin2011}. Many of these arise from the unique behavior of its low-energy electrons, which resemble relativistic particles \cite{novoselov2005}. For example, charge carriers in graphene can lead to phenomena present in quantum electrodynamics by the massless fermions, but at lower speeds \cite{araki2012}. Importantly, this quasiparticle behavior of electrons is inherently related to the triangular sublattice arrangement of atoms within the hexagonal structure. In this context, one of the most fundamental aspects of these charge carriers is their chiral character defined through specific pseudospin vector related to the two-component wave function of the Dirac quasiparticles. As such, this chiral nature is responsible for many extraordinary properties of graphene, including Klein tunneling \cite{katsnelson2006_1}, the half-integer quantum Hall effect \cite{jacak2012}, or the absence of backscattering at $p$-$n$ junctions \cite{katsnelson2006_2}.

It is then widely regarded that the intrinsic chirality in pristine graphene is intriguing on its own, giving rise to a broad range of potential semimetallic applications. Yet, graphene is also expected to be a promising condensed matter host for various other phases when the discussed chiral symmetry is broken. The bond density wave is one such representative phase that couples electrons from opposite valleys, breaks translational symmetry, and introduces alternating bond order in graphene \cite{qu2022}. This mechanism constitutes a two-dimensional analogy to the Peierls instability and is commonly known as the Kekul{\'e} distortion. In practice, it can be induced in graphene by adatoms or an appropriate substrate. An example of the former is the decoration of graphene with lithium or hydrogen \cite{bao2021, guan2024}, while the latter can be realized via coupling to topological insulators \cite{lin2017}. In both cases, the result is the generation of a {\it fermionic mass} and the opening of a band gap at the Dirac point in graphene. However, while the second symmetry-breaking approach is suggested to yield a semiconducting phase of graphene, in the first case adatoms lift the Fermi energy into the conduction band, inducing a metallic state.

Due to the nature of Kekul{\' e} distortion, the associated phases of graphene are also expected to present strongly altered phonon dynamics, leading to the increased role of interactions between lattice vibrations and now {\it mass} electrons. Indeed, such graphene structures have been found to exhibit intriguing phonon-mediated effects, including topological transport \cite{wu2016, liu2017, garcia2024}, and most notably superconductivity \cite{profeta2012, ludbrook2015, szczesniak2019}. However, despite this interest, the electron-phonon correlations remain only partially explored. Of particular interest is the local insight, as indicated by the spatial character of discussed distortion \cite{sorella2018, eom2020}, its potential of being driven by the electron-phonon interaction \cite{zhang2021} and the role in shaping or reinforcing related quantum phase known as the Kekul{\' e} valence bond order \cite{costa2024}. Unfortunately, most studies in this respect focus primarily on the uniform or mean-field properties \cite{costa2024, otsuka2024}. This is to say, a deeper understanding is hindered, as the influence of local lattice deformations tends to be averaged out. As a consequence, the in-plane dynamics are often heavily approximated or even neglected when discussing phonon-mediated effects in the presence of Kekul{\' e} bond order. This is particularly apparent in studies on superconductivity, which tend to focus on out-of-plane states only \cite{krok2023}, although the in-plane interactions may still notably modify the overall electron-phonon coupling \cite{kaloni2013, pesic2014, szczesniak2023}.

According to the above, motivation arises to provide quantitative framework for the local analysis of electron-phonon coupling in Kekul{\' e}-ordered graphene. Here it is argued that this can be done effectively by recognizing that the signatures of the Kekul{\' e} bond order should be particularly well reflected in the behavior of the phonon modes which involve in-plane bond-stretching vibrations of carbon atoms, {\it e.g.} the $E_{2g}$ mode. In other words the analysis is suggested to concentrate on the phonons that are highly sensitive to the changes in bond lengths caused by the discussed distortion. These are the phonons, associated with the bond modulations, that are also expected to have pronounced impact on the electron-phonon coupling in graphene \cite{castro2007}. When following \cite{cappelluti2012} in spirit, the described effect can be quantified within the modulation of electronic coupling between neighboring carbon atoms, by recalling the concept of deformation potential \cite{khan1984}. In details, it can be stated that the magnitude of the electron-phonon coupling is meant to depend strongly on the shift in electronic energy levels induced by a frozen-phonon lattice distortion. With this in mind, it is possible to discuss in detail the electron-phonon coupling at the local level. this approach not only provides novel and instructive insight into the interactions of interest, but also lays the foundation for discussing related phenomena, such as the previously mentioned superconductivity.

%%%%%%%%%%%%%%%%%%%%%%%%%%%%%%%%%%%%%%%%%%%%%%%%%%%%%%%%%%%%%%%
\section{Methodology}
%%%%%%%%%%%%%%%%%%%%%%%%%%%%%%%%%%%%%%%%%%%%%%%%%%%%%%%%%%%%%%%

\begin{figure}[ht!]
\includegraphics[width=\columnwidth]{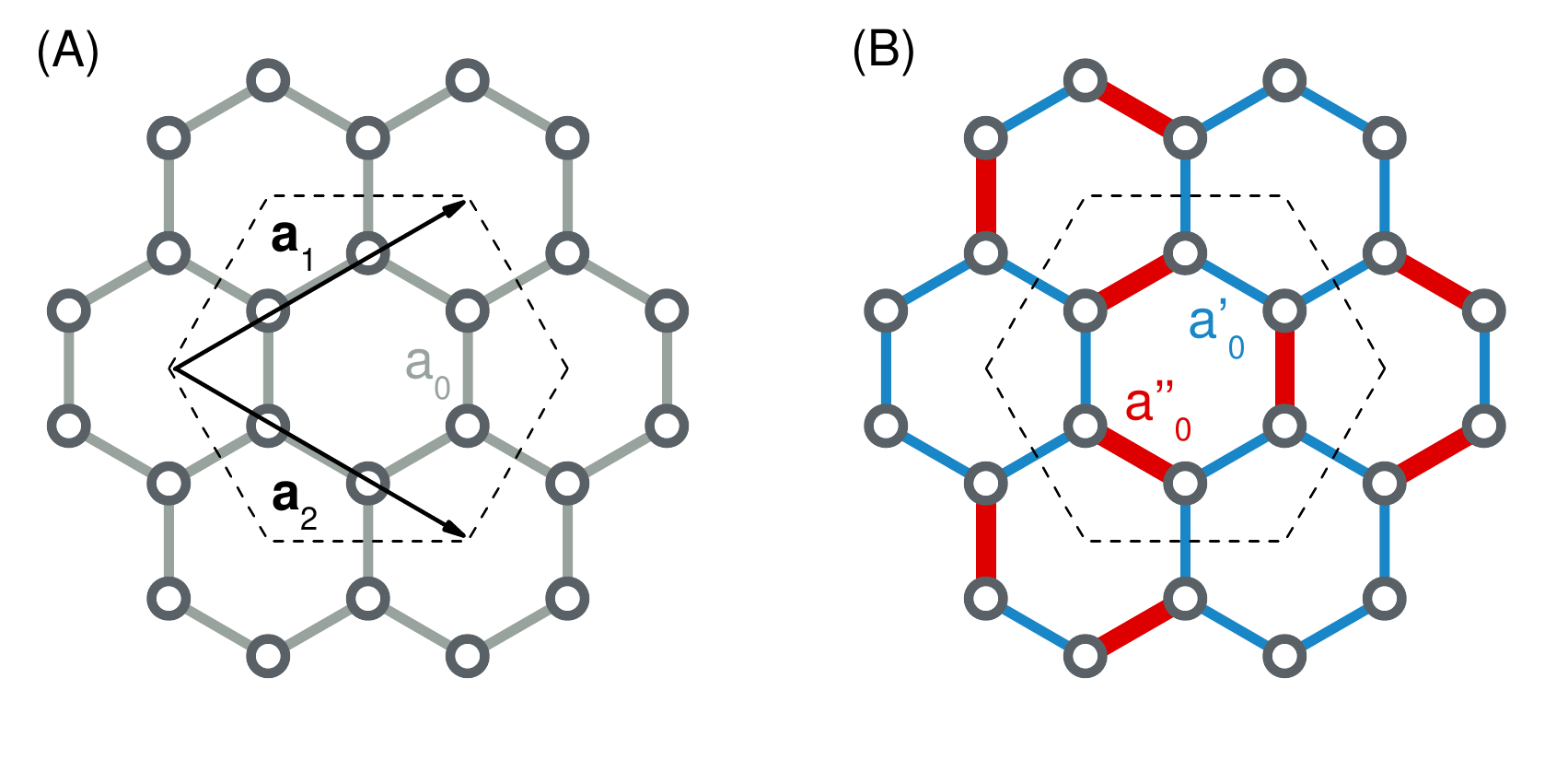}
\caption{The schematic representation of pristine graphene (A) and graphene with Kekul{\' e} bond order (B). The unaltered bonds (length $a$) are marked grey, while the red and blue colors correspond to the shorter (length $a'$) two-third and longer (length $a''$) one-third distorted bonds, respectively. The $\left(\sqrt{3} \times \sqrt{3} \right) {\rm R} 30^{o}$ supercell is depicted in green along with the the corresponding basis/unit lattice vectors.}
\label{fig01}
\end{figure}

To induce chiral symmetry breaking in graphene through Kekul{\' e} distortion, it is convenient to consider $\left(\sqrt{3} \times \sqrt{3} \right) {\rm R} 30^{o}$ supercell arrangement of carbon atoms. This configuration uniquely captures the intrisic pattern of the discussed bond distortion and results in the unit cell that consists of six atomic sites. In this case, the corresponding lattice vector that enables reconstruction of the infinite graphene layer reads:
\begin{equation}
\label{eq1}
\mathbf{R} = \alpha \mathbf{a}_{1} + \beta \mathbf{a}_{2},
\end{equation}
where $\mathbf{a}_{1}=(\sqrt{3}/2, 1/2)a$ and $\mathbf{a}_{2}=(\sqrt{3}/2, -1/2)a$ are the unit lattice vectors, with the lattice parameter $a=3a_0$ defined in terms of the nearest neighbor distance $a_0=1.42$ \AA. Given this, $\alpha$ and $\beta$ take on integer values. For graphical reference please see Fig. \ref{fig01} (A).
According to the above, the modulation of the nearest neighbor distance appears to underlie the Kekulé distortion. This is formalized here by recalling the fact that the chiral symmetry in graphene refers to its low-energy electronic properties, so it can be described with respect to the $p_{z}$ electrons only. These are the excitations with strong coupling between pseudospin and momentum, responsible for forming $\pi$-type valence and conduction bands that dominate electronic structure near the Dirac points. Hence, assuming that the Kekulé bond order leads to a gapped structure, the interplay between discussed structural and electronic features is effectively captured by the following Hamiltonian:
\begin{eqnarray}
\label{eq2}
\nonumber
H &=& \sum_{i} \epsilon a^{\dagger}_{i} c_{i} - \sum_{\left< i,j \right>} t_{0} c^{\dagger}_{i} c_{j} - \sum_{\left< \left< i,j \right> \right>} t_{1} c^{\dagger}_{i} c_{j} - \\
&-& \sum_{\left< \left< \left< i,j \right> \right> \right>} t_{2} c^{\dagger}_{i} c_{j},
\end{eqnarray}
where $\epsilon$ is the energy of an electron in the $p_{z}$ orbital located on the $i$-th carbon site, while $t_{n}$ is the $n$-th nearest neighbor coupling integral between two $i$-th and $j$-th carbon sites, with $n \in \left< 0,2 \right>$. Thus, $c^{\dagger}_{i}$ ($c_{j}$) create (annihilate) electrons in the $p_{z}$ state for an appropriate pair of carbon atoms at the distance $a_0$, $a_{1}=a_{0}\sqrt{3}$ and $a_{2}=2a_{0}$ for the first, second and third nearest nearest neighbors, respectively. In this form, Eq. (\ref{eq2}) balances computational efficiency and predictive accuracy, allowing for refined representation of the Kekulé-induced dispersion due to the broken electron-hole symmetry.

In the framework of Eq. (\ref{eq2}), the Kekul{\' e} distortion can be now formally introduced as:
\begin{equation}
\label{eq3}
t_{0} = 2/3 t_{0}'  + 1/3 t_{0}'',
\end{equation}
where $t_{0}'$ and $t_{0}''$ are the unbalanced couplings for the two alternating bond types that compose the structural pattern of interest. In detail, the $t_{0}'$ couplings are associated with the $a'_{0}$ bonds, which are shorter than $a_{0}$ and account for two-thirds of all distorted bonds, while the $t_{0}''$ couplings correspond to the longer $a''_{0}$ bonds, making up for the remaining one-third. The schematic representation of graphene with Kekul{\' e} distortion is presented in details in Fig. \ref{fig01} (B). Note that, as a result of the introduced distortion, the distances beyond the first nearest neighbor approximation also alternate and are given by:
\begin{equation}
\label{eq4}
a'_{1}=a_{0}'\sqrt{3},
\end{equation}
\begin{equation}
\label{eq4}
a''_{1}=\sqrt{(a_{0}')^{2}+a_{0}'a_{0}''+(a_{0}'')^{2}},
\end{equation}
and,
\begin{equation}
\label{eq5}
a'_{2}=2a'_{0},
\end{equation}
\begin{equation}
\label{eq5}
a''_{2}=(a''_{0})^{2}+(a'_{1})^{2}/\sqrt{(a''_{0})^{2}+(a'_{1})^{2}},
\end{equation}
where $a'_{1}(a''_{1})$ and $a'_{2}(a''_{2})$ represent the second and third nearest neighbor distances, respectively, for atom pairs located within (between) unit cells of the distorted graphene structure given in Fig. \ref{fig01} (B). It means that the $t_{1}$, and $t_{2}$ coupling integrals of Eq. (\ref{eq2}) become modified with respect to the changes introduced via Eq. (\ref{eq3}). This is to say, Eq. (\ref{eq2}) fully incorporates the influence of the Kekul{\' e} distortion, although this distortion is introduced with respect to the first nearest neighbor couplings only. In this context, the magnitude of symmetry breaking can be captured by the energy gap width, expressed in terms of the unbalanced couplings as:
\begin{equation}
\label{eq6}
\Delta = 2|t_{0}''-t_{0}'|,
\end{equation}
in the nearest neighbor approximation.

%%%%%%%%%%%%%%%%%%%%%%%%%%%%%%%%%%%%%%%%%%%%%%%%%%%%%%%%%%%%%%%
\section{Electronic dispersion}
%%%%%%%%%%%%%%%%%%%%%%%%%%%%%%%%%%%%%%%%%%%%%%%%%%%%%%%%%%%%%%%

The energy and coupling parameters exploited in the present paper are derived from the model by Gr{\"u}neis {\it et al.} \cite{gruneis2008}, which can be considered universal for the purpose of describing electronic structure of graphene, especially in terms of its low-energy properties. However, to allow incorporation of the phonon effects in question, the explicit dependence on the bond length $l$ for the latter parameters has to be added. In general, beyond the equilibrium distance, the renormalization of the electronic coupling in graphene can be given as \cite{pereira2009}:
\begin{equation}
\label{eq7}
t=t_{0}e^{\eta(l / a_{0}-1)},
\end{equation}
where $t_{0}$ is the equilibrium coupling between two carbon sites at the distance $a_{0}$ and $\eta$ denotes the fall-off coefficient, interpreted as the phonon mode softening/hardening rate \cite{naumis2017}. Here, $\eta$ is calculated by fitting Eqs. (\ref{eq7}) to the respective set of parameters by Gr{\"u}neis {\it et al.} \cite{gruneis2008}, using the Levenberg-Marquardt algorithm \cite{levenberg1944, marquardt1963}. That means fitting procedure is conducted with respect to not only the nearest ($t_{0}$) but also distant neighbor terms ($t_{1}$, and $t_{2}$). This ensures that $\eta$ effectively accounts for the high-order contributions related to the electron-hole and structural asymmetries, familiar in the Kekul{\' e} scenario. For the parameters corresponding to the local density approximation the $|\eta|$ value obtained here is $\sim 2.13$, suggesting notable reduction from the previous nearest neighbor fits that yield this rate to be of the order of 3 or more for graphene \cite{castro2007, ribeiro2009, pereira2009, kundu2009}. However, this estimate is much closer to the $\eta$ value for the critical bond modulation required to open a band gap in graphene \cite{botello-mendez2018}. Consequently, the observed reduction appears to arise from the smaller and more realistic bond modulation necessary to induce a gap in the third nearest neighbor model, as compared to the nearest neighbor approximation. Note that this nontrivial insight follows some earlier studies \cite{botello-mendez2018, reich2002}. 

\begin{figure}[ht!]
\includegraphics[width=\columnwidth]{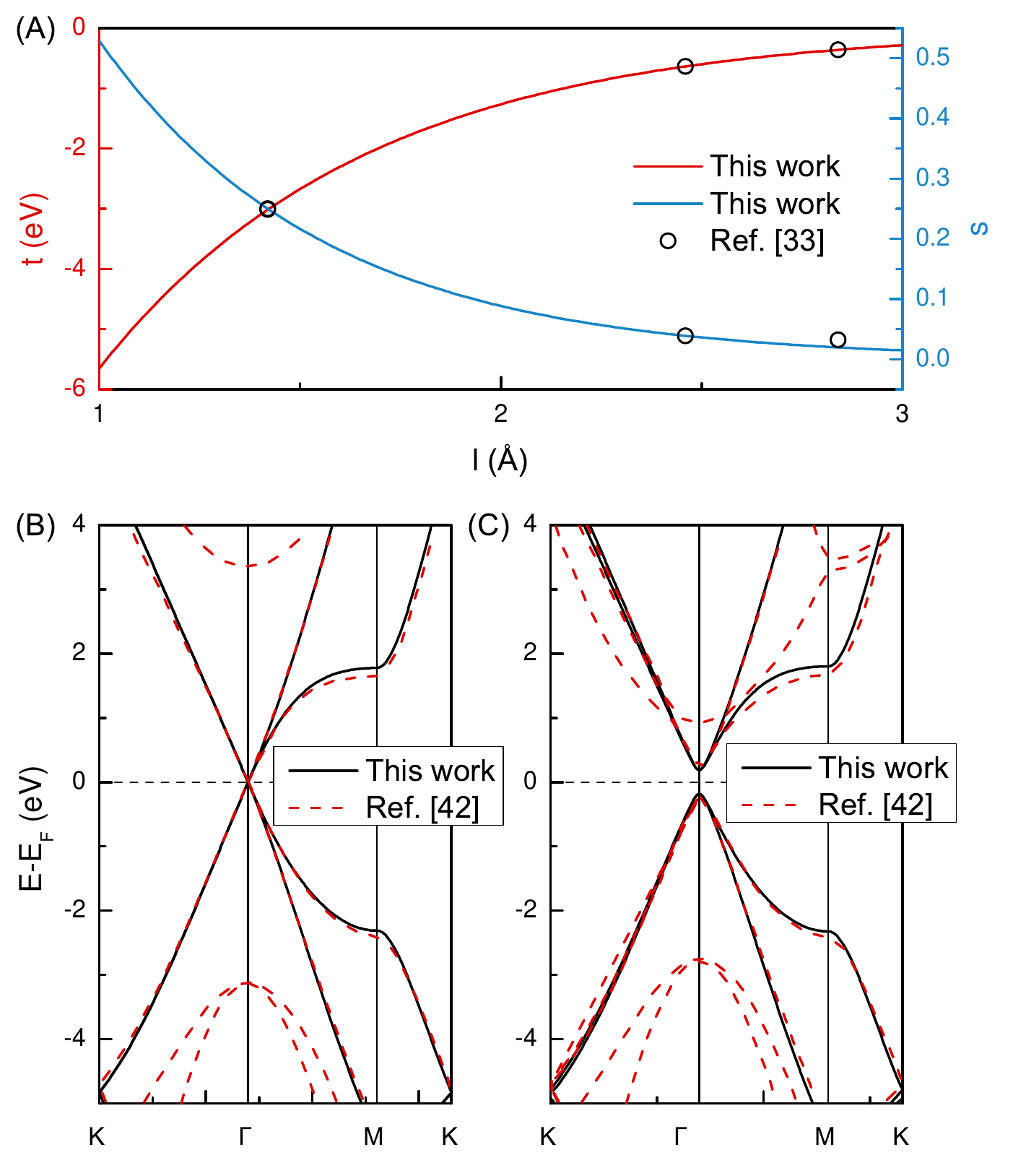}
\caption{The electronic coupling and overlap parameters as a function of distance between neighboring carbon atoms (A). The solid lines correspond to the results obtained via distance dependent relations, whereas open circles are the equilibrium estimates at the distance $a_{0}$ derived from \cite{gruneis2008}. The electronic band structure of pristine (B) and distorted graphene (C) for the $\left(\sqrt{3} \times \sqrt{3} \right) {\rm R} 30^{o}$ supercell. The results obtained by using the effective model are given by solid lines while dashed lines mark results of the density function theory calculations assuming local density approximation, given in \cite{farjam2009}. The Fermi energy is set at the zero reference level.}
\label{fig02}
\end{figure}

In Fig. \ref{fig02} (A) the distance dependence of the renormalized electronic coupling is presented, as obtained from the Eq. (\ref{eq7}). The electronic coupling decay exponentially, in agreement with the simultaneous reduction of the overlap between involved orbitals. The latter is also depicted, being derived for the distance dependent relation, analogous to Eq. (\ref{eq7}): $s=s_{0}\exp{[\eta'(l / a_{0}-1)]}$, where $s_{0}=0.2499$ \cite{gruneis2008} and $|\eta'|=2.5359$. This is the expected behavior which is suggested by the two-center approximation \cite{slater1954}. For comparison purposes the equilibrium values of electronic couplings at the distance $a_{0}$ are marked by the open circles. These are directly adopted from \cite{gruneis2008} and exhibit an almost perfect match with the results of Eq. (\ref{eq7}), further reinforcing its validity.

The developed effective model is additionally verified through calculations of the electronic band structures and their comparison with the results obtained from more sophisticated methods. In particular, two band structures for the pristine and distorted graphene are considered, assuming the $\left(\sqrt{3} \times \sqrt{3} \right) {\rm R} 30^{o}$ supercell. While the gapless case is self-explanatory, the distorted structure is calculated for the scenario where the band gap induced by the Kekul{\' e} bond order is $\sim 0.4$ eV. Note that this is a representative empirical value corresponding to the lithium-decorated graphene \cite{bao2021}. For both systems, the reference data is assumed after \cite{farjam2009}, where it was obtained from the density functional theory calculations in the local density approximation. In order to improve matching between prediction of the effective model and the density function theory, the overlap parameters are also considered when deriving dispersion relations from Eq. (\ref{eq2}). As already mentioned, these parameters are treated distance dependent, similarly to the electronic couplings described by the Eq. (\ref{eq7}). Analogous fitting procedure for the overlap parameters is performed. Their respective behavior is presented in Fig. \ref{fig02} (B). Once again, obtained results are compared with the equilibrium values at the distance $a_{0}$, assumed after \cite{gruneis2008}, confirming relatively high accuracy of the distance dependent formula.

The calculated band structures of pristine and distorted graphene structures are depicted in Figs. \ref{fig02} (B) and (C), respectively. The effective model results are given by the solid lines and compared with the predictions of the density function theory marked by the dots. The low-energy features appear to be well reproduced in the framework of Eq. (\ref{eq2}), both for the gapless and distorted case, showing expected transition from the linear to parabolic dispersion. Despite the fact that matching between two considered sets of results slightly decreases, as we move away from the Fermi energy, the characteristics of the high-energy regions are still conserved. For example the threefold degeneracy at the K point is lifted, additionally suggesting accuracy of the assumed relation for the unbalanced couplings in Eq. (\ref{eq3}).

Clearly, the effective model does not account for all the bands due to its inherent limitations. Nonetheless, the reference data is provided even for bands beyond the $\pi$-type approximation, to emphasize the significance of the low-energy dispersion. These bands, primarily located around the $\Gamma$ point, have minimal impact on the physics near the Fermi energy, {\it e.g.}, in terms of the phonon-assisted transport processes, and remain largely unaffected by the introduction of the bond density wave order. Still, it is important to note that when decoration with adatoms is used to break chiral symmetry in graphene, the additional low-energy states (the interlayer states) can be introduced. Here such state, arising from the $s$ orbitals of the lithium atoms, is visible just above the band gap, in Fig. \ref{fig02} (C). This state is responsible for rising Fermi energy into the conduction band and complementing $\pi$-type bands contribution to the van Hove singularity. Although important for phenomena such as conventional superconductivity \cite{profeta2012, ludbrook2015, szczesniak2019}, due to its positive contribution to the total electron-phonon coupling, it does not solely governs the interactions between electron and phonons but rather allows promotion of the out-of-plane vibrations. Consequently, it is evident that a thorough analysis of the in-plane coupling is essential for a complete understanding of the effects induced by the Kekul{\' e} distortion.

%%%%%%%%%%%%%%%%%%%%%%%%%%%%%%%%%%%%%%%%%%%%%%%%%%%%%%%%%%%%%%%
\section{Local electron-phonon coupling}
%%%%%%%%%%%%%%%%%%%%%%%%%%%%%%%%%%%%%%%%%%%%%%%%%%%%%%%%%%%%%%%

The fact that in-plane phonons indeed induce energy band splitting in graphene \cite{samsonidze2007}, as suggested by the deformation potential concept, implies that the local electron-phonon term may be quantified with respect to the modulations of electronic coupling caused by the carbon atom displacements. Here, this effect is uniquely captured by the bond-resolved electron-phonon coupling, expressed as \cite{castro2007, cappelluti2012}:
\begin{equation}
\label{eq8}
\alpha=dt/dl,
\end{equation}
and considered within the frozen-phonon approach. In this framework, the displacement mimics the in-plane $E_{2g}$ phonon mode that influences electronic dispersion according to the response of $t$. Assuming that $t$ follows Eq. (\ref{eq7}), the Eq. (\ref{eq8}) allows to expect that $\alpha$ and $t$ are linearly dependent. Note that this is not a universal property, as it does not hold for all the potential forms of $t$, {\it e.g.}, when considering Harrison flyleaf expression that suggests $t\sim1/l^{2}$. This is to say, it can be argued that the linear character of Eq. (\ref{eq8}) is a results of the fact that Eq. (\ref{eq7}) appears to be inherently valid for Kekul{\' e} scenario, as shown in the previous section.

\begin{figure}[hb!]
\includegraphics[width=\columnwidth]{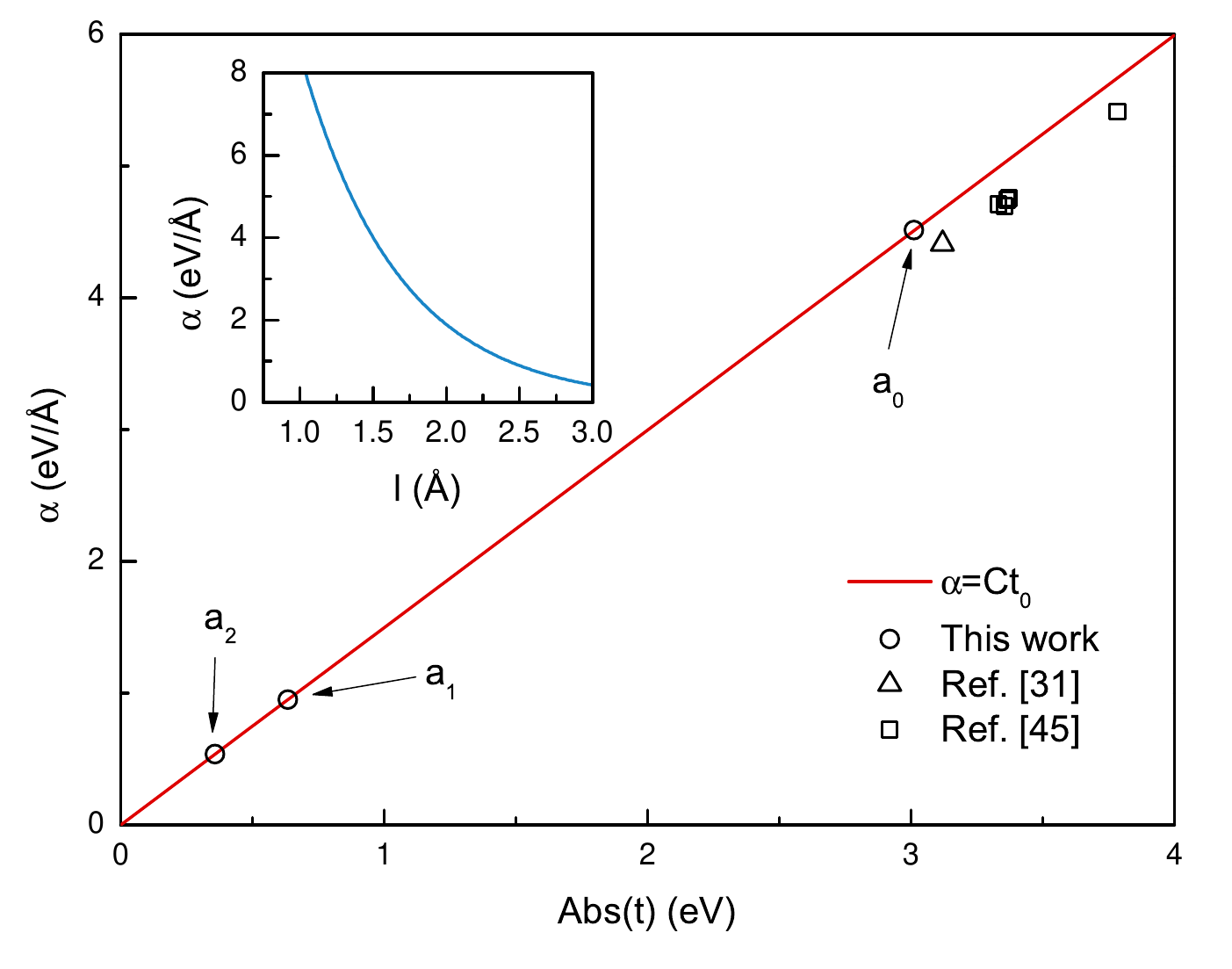}
\caption{The local electron-phonon coupling against absolute value of the electronic coupling. Corresponding pair estimates derived from various approaches are depicted with open circles (this work), triangles (Ref. [47]) and squares (Ref. [48]). The inset presents dependence of the local electron-phonon coupling on the lattice distortion.}
\label{fig03}
\end{figure}

In Fig. (\ref{fig03}) the behavior of the electron-phonon coupling in distorted graphene is presented as a function of the electronic coupling given by Eq. (\ref{eq7}). Note that for convenience the absolute value of $t$ is considered. It is clear that $\alpha$ indeed scales linearly with $t$, following earlier argumentation. In details, these parameters are correlated as:
\begin{equation}
\label{eq10}
\alpha=Ct,
\end{equation}
where $C=1.49817$ \AA$^{-1}$. Interestingly, this finding is in close agreement with earlier studies on graphene systems that suggest such behavior for various phonon modes independently \cite{lazzeri2008, cappelluti2012}. For comparison, these are depicted in Fig. (\ref{fig03}) by the open triangles and squares, almost perfectly matching the proposed here scaling trend. The small deviations are likely due to the fact that the semi-empirical part of analysis in \cite{lazzeri2008, cappelluti2012} was conducted within the nearest neighbor regime. Moreover, the values presented in \cite{cappelluti2012} were optimized specifically for bilayer graphene, whereas \cite{lazzeri2008} reported estimates primarily derived from methods beyond the local density approximation. Still, the obtained here results not only confirm that the linear correlation is independent of the $t$ value, as initially suggested in \cite{cappelluti2012}, but also that it does not depend on the used approximation and specifically follows the functional form of Eq. (\ref{eq7}).

As such, the above allows to address another fundamental aspect, namely how the electron-phonon interaction changes with the interatomic distance in graphene. This dependence is presented in the inset of Fig. (\ref{fig03}), suggesting that the electron-phonon coupling, due to the in-plane phonons, has potential to rise quite notably from the equilibrium distance $a_{0}$, along with the decrease of the distance between carbon atoms. The corresponding change rate is $\left|d\alpha/dl\right|=9.92$ eV/\AA$^{2}$ for $l \in \left< a_{0}-\delta, a_{0} \right>$ when $\delta=0.5$ \AA. Interestingly enough, the opposite appears to be less effective, as $\alpha$ reduces slower with the increasing distance and begins to saturate around 3 \AA, which is the doubled equilibrium distance $a_{0}$. In this case the change rate is $\left|d\alpha/dl\right|=4.72$ eV/\AA$^{2}$ assuming $l \in \left<  a_{0}, a_{0}+\delta \right>$. The observed behavior reflects the fact that the electronic coupling is far more sensitive to orbital overlap at shorter distances and that overlap falls off more slowly with increasing distance. In other words, in-plane local electron-phonon interaction is not uniform across both regions and contributes less to the total coupling with the increased interatomic distance. In fact, the latter partially explains previously observed decrease in the electron-phonon coupling when the tensile equibiaxial strain is applied to the lithium-decorated graphene \cite{pesic2014}, implying strong interplay between in-plane and out-of-plane vibrations in this system. Note that this is not what is expected from the influence of strain alone, as expected for the 2D materials \cite{zhou2015, wan2013}. Still the observed effects may be beneficial when aiming at the enhancement of the electron-phonon coupling in graphene. However, for this purpose, the obtained results indicate that gradient between two distinguished regions should be relatively large. This would suggest need for an external factor that will allow to engineer bonds in Kekul{\' e}-ordered graphene in a non-uniform manner.

\begin{figure}[ht!]
\includegraphics[width=\columnwidth]{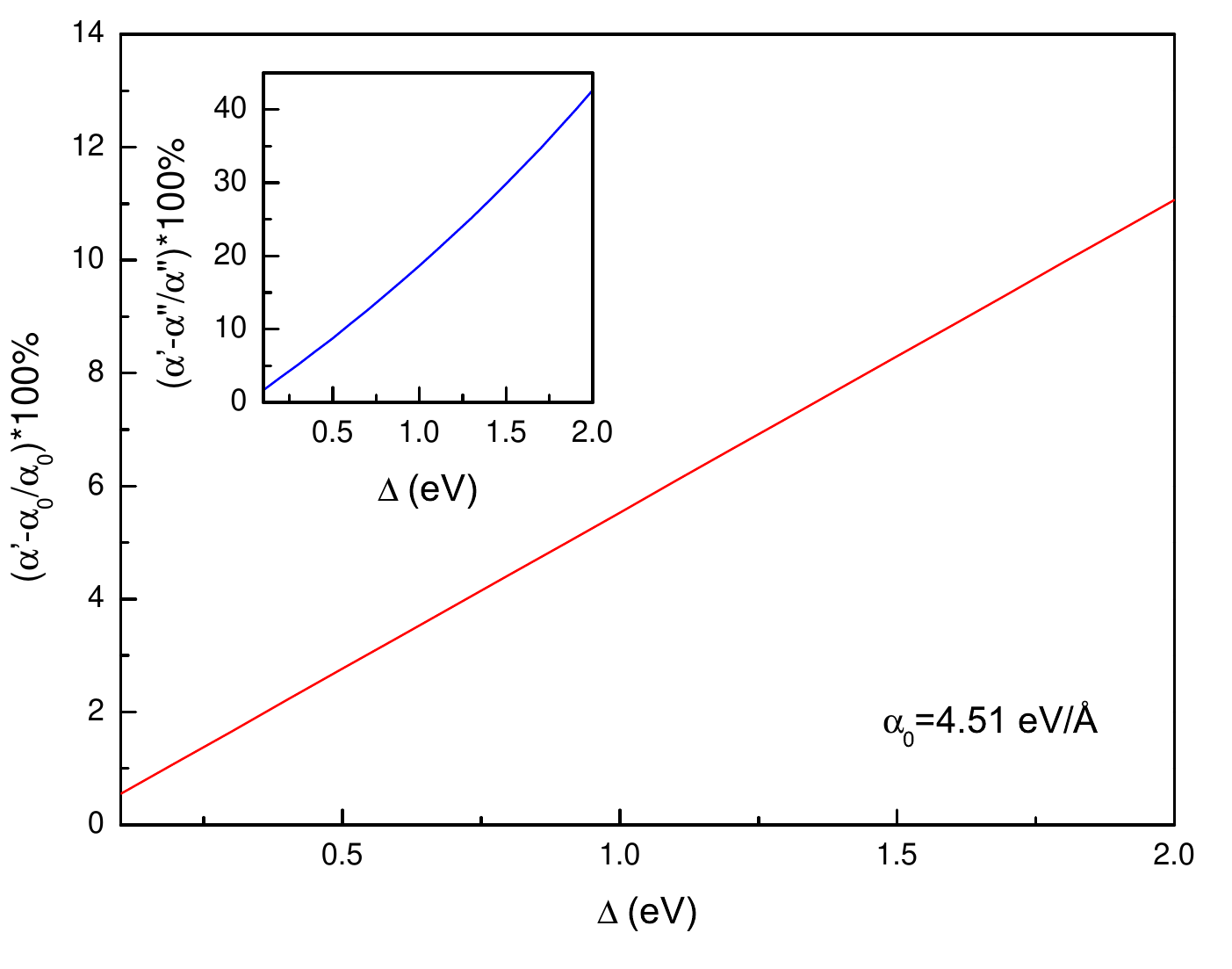}
\caption{The percentage increase of the local electron-phonon coupling for the two-third bonds against its equilibrium value, as calculated for the graphene with Kekul{\' e} bond order. The inset depicts percentage increase for the same parameter but against the local electron-phonon coupling for the one-third bonds.}
\label{fig04}
\end{figure}

To discuss in more detail the observed anisotropy of the local electron-phonon coupling in disordered graphene, and to estimate the potential for engineering it through Kekul{\' e} distortion, it is instructive to conduct the analysis with respect to the band gap size. This way it should be possible to establish realistic limits for the electron-phonon coupling according to the empirical observable. For that reason, several distorted graphene structures are considered, which are characterized by the band gap values ranging from 0.1 eV to 2 eV. Thus, the conducted analysis encapsulates exemplary case of lithium-decorated graphene with band gap of $\sim 0.4$ \cite{bao2021} but also some extreme instances considered in the literature \cite{khoa2019, park2015}. This not only enables to examine observed effects across a broader range but also to trace the overall behavior of the electron-phonon coupling.

From the global perspective, the most influential in terms of tailoring the electron-phonon coupling in graphene appears to be bonds that become shorter as a result of the Kekul{\' e} distortion. As previously noted, these bonds constitute the majority, exactly two-thirds, of all distorted bonds. Hence, the change in local electron-phonon coupling, induced by the bond shortening, is expected to provide a key initial insight into the discussed effects. In Fig. \ref{fig04}, the relative variation in strength of the electron-phonon correlations for the two-thirds of the bonds ($\alpha'$) is expressed as a percentage change from the coupling at equilibrium distance ($\alpha_{0}$): $(\alpha'-\alpha_{0}/\alpha_{0})\times100\%$. The resulting values are plotted as a function of the band gap size ($\Delta$). It can be observed that the discussed local electron-phonon coupling strengthens linearly with the band gap. The maximum coupling is reach at the band gap equal of 2 eV, exceeding the equilibrium value by more than 10 \%. At the same time, the lattice becomes locally stiffer as the band gap widens. This indicates that the two trends strongly compete in Kekul{\' e}-ordered graphene, a fact that is particularly important in light of the potential superconducting phase that may emerge in the discussed structure. In particular, the results suggest that in order to enhance the superconducting state, electron-phonon renormalization must dominate over the opposing mechanical effects. In other words, the stabilization of such a superconducting phase likely requires the system to be in the strong coupling regime. This partially explains why superconductivity characterized by the strong interactions between electrons and phonons was observed in the lithium-decorated graphene and its derivatives \cite{szczesniak2019, szczesniak2023}. However, at the same time this analysis suggest increased phonon energies \cite{pesic2014} that may contribute to the suggested non-adiabatic character of such phase \cite{szczesniak2019, szczesniak2023}.

Of course, even bigger percentage change in the electron-phonon coupling is expected when considering interplay between terms for the shorter two-third and longer one-third bonds. In the inset of Fig. \ref{fig04}, the corresponding relation is plotted against the band gap size, expressed relative to the coupling for the longer bonds ($\alpha''$) as: $(\alpha'-\alpha''/\alpha'')\times100\%$. Once again, the ratio increases with the band gap, showing a slight deviation from perfect linearity, and reaches a maximum relative change exceeding 40 \%. This is to say, the two-thirds of bonds give rise to centers of locally enhanced electron-phonon coupling, periodically distributed across the graphene sheet. Naturally, the impact of these centers on the total electron-phonon coupling is compensated to some extend by the contributions coming from the longer one-third bonds. Nevertheless, it is important to note here that the identified regions should still be considered when examining phenomena that are highly sensitive to local variations in material properties.

%%%%%%%%%%%%%%%%%%%%%%%%%%%%%%%%%%%%%%%%%%%%%%%%%%%%%%%%%%%%%%%
\section{Conclusions and perspectives}
%%%%%%%%%%%%%%%%%%%%%%%%%%%%%%%%%%%%%%%%%%%%%%%%%%%%%%%%%%%%%%%

In conclusion, the present study demonstrates that the behavior of local electron-phonon coupling in graphene is strongly correlated with the structural modulations, as introduced by the Kekul{\' e} bond order. In particular, the local strength of the electron-phonon interaction is found to scale linearly with the electronic coupling, which was optimized to capture the distortion-induced modifications in the electronic dispersion of graphene. In the context of previous studies, these findings contribute to a unified picture of how electron-phonon coupling in graphene fluctuates in response to local structural changes, revealing a universal trend that can be readily applied in future research. For instance, it should be possible to use this well-defined relationship in experiment or with more sophisticated methods not only to estimate the magnitude of electron-phonon correlations based on empirical or theoretically-precise electronic coupling, but also to qualitatively track its variation under external influences of strain, doping, substrate or fields. Furthermore, due to its simplicity, the relation holds promise for machine learning applications, where it could serve as a constraint in the training and validation of models designed to analyze electron-phonon interactions in graphene.

The significance of these findings is underscored by their implications for phonon-mediated phenomena, providing new insights into the underlying mechanisms and suggesting avenues for controlled engineering. Of particular importance are the observations made in terms of the superconducting phase that induces in the Kekul{\' e}-ordered graphene. Specifically, the identified local behavior of the electron-phonon interaction partially explains strong coupling, or even non-adiabatic, character of such state and proves capability of the in-plane strain to enhance it. However, further inspection reveals that introduced distortions yield centers of locally increased electron-phonon coupling that may also play role in optimizing properties of interest. In this manner, it can be argued that graphene with Kekul{\' e} bond order should be viewed as a material with spatially non-uniform pairing strength that requires specific methods to tune the electron-phonon interactions. By analogy to granular superconductors, one may consider strategies aimed at modulating the resonance between enhancement centers toward most efficient superconducting state the coherently {\it percolates} through the material. On the the other hand, these regions can act as a quasi-periodic phonon scattering centers, affecting resistivity and leading to highly anisotropic or non-linear transport behavior.

To this end, the behavior of electron-phonon coupling in Kekul{\' e}-ordered graphene appears to the more complex than previously anticipated, thanks to the presented here insight. In particular, the discussed correlations show rich physics and interplay at the local level that explains some of their peculiar aspects, hitherto not fully addressed. The conducted analysis implies that any attempt to engineer electron-phonon coupling in graphene with Kekul{' e} bond order should account for the strong local anisotropy of this interaction, rather than treating it as a uniformly distributed characteristic. In fact, the presented findings may provide a new paradigm for such efforts. For this purpose, the effective model presented here may serve as a useful foundation for further studies.

%%%%%%%%%%%%%%%%%%%%%%%%%%%%%%%%%%%%%%%%%%%%%%%%%%%%%%%%%%%
\bibliographystyle{apsrev}
\bibliography{manuscript}

\begin{thebibliography}{48}
\expandafter\ifx\csname natexlab\endcsname\relax\def\natexlab#1{#1}\fi
\expandafter\ifx\csname bibnamefont\endcsname\relax
  \def\bibnamefont#1{#1}\fi
\expandafter\ifx\csname bibfnamefont\endcsname\relax
  \def\bibfnamefont#1{#1}\fi
\expandafter\ifx\csname citenamefont\endcsname\relax
  \def\citenamefont#1{#1}\fi
\expandafter\ifx\csname url\endcsname\relax
  \def\url#1{\texttt{#1}}\fi
\expandafter\ifx\csname urlprefix\endcsname\relax\def\urlprefix{URL }\fi
\providecommand{\bibinfo}[2]{#2}
\providecommand{\eprint}[2][]{\url{#2}}

\bibitem[{\citenamefont{{Castro Neto} et~al.}(2009)\citenamefont{{Castro Neto},
  Guinea, Peres, Novoselov, and Geim}}]{castro2009}
\bibinfo{author}{\bibfnamefont{A.~H.} \bibnamefont{{Castro Neto}}},
  \bibinfo{author}{\bibfnamefont{F.}~\bibnamefont{Guinea}},
  \bibinfo{author}{\bibfnamefont{N.~M.~R.} \bibnamefont{Peres}},
  \bibinfo{author}{\bibfnamefont{K.}~\bibnamefont{Novoselov}},
  \bibnamefont{and} \bibinfo{author}{\bibfnamefont{A.~K.} \bibnamefont{Geim}},
  \bibinfo{journal}{Reviews of Modern Physics} \textbf{\bibinfo{volume}{81}},
  \bibinfo{pages}{109} (\bibinfo{year}{2009}).

\bibitem[{\citenamefont{Bonaccorso et~al.}(2010)\citenamefont{Bonaccorso, Sun,
  Hasan, and Ferrari}}]{bonaccorso2010}
\bibinfo{author}{\bibfnamefont{F.}~\bibnamefont{Bonaccorso}},
  \bibinfo{author}{\bibfnamefont{Z.}~\bibnamefont{Sun}},
  \bibinfo{author}{\bibfnamefont{T.}~\bibnamefont{Hasan}}, \bibnamefont{and}
  \bibinfo{author}{\bibfnamefont{A.~C.} \bibnamefont{Ferrari}},
  \bibinfo{journal}{Nature Photonics} \textbf{\bibinfo{volume}{4}},
  \bibinfo{pages}{611} (\bibinfo{year}{2010}).

\bibitem[{\citenamefont{Lee et~al.}(2008)\citenamefont{Lee, Wei, Kysar, and
  Hone}}]{lee2008}
\bibinfo{author}{\bibfnamefont{C.}~\bibnamefont{Lee}},
  \bibinfo{author}{\bibfnamefont{X.}~\bibnamefont{Wei}},
  \bibinfo{author}{\bibfnamefont{J.~W.} \bibnamefont{Kysar}}, \bibnamefont{and}
  \bibinfo{author}{\bibfnamefont{J.}~\bibnamefont{Hone}},
  \bibinfo{journal}{Science} \textbf{\bibinfo{volume}{321}},
  \bibinfo{pages}{385} (\bibinfo{year}{2008}).

\bibitem[{\citenamefont{Balandin}(2011)}]{balandin2011}
\bibinfo{author}{\bibfnamefont{A.~A.} \bibnamefont{Balandin}},
  \bibinfo{journal}{Nature Materials} \textbf{\bibinfo{volume}{10}},
  \bibinfo{pages}{569} (\bibinfo{year}{2011}).

\bibitem[{\citenamefont{Novoselov et~al.}(2005)\citenamefont{Novoselov, Geim,
  Morozov, Jiang, Katsnelson, Grigorieva, Dubonos, and Firsov}}]{novoselov2005}
\bibinfo{author}{\bibfnamefont{K.~S.} \bibnamefont{Novoselov}},
  \bibinfo{author}{\bibfnamefont{A.~K.} \bibnamefont{Geim}},
  \bibinfo{author}{\bibfnamefont{S.~V.} \bibnamefont{Morozov}},
  \bibinfo{author}{\bibfnamefont{D.}~\bibnamefont{Jiang}},
  \bibinfo{author}{\bibfnamefont{M.~I.} \bibnamefont{Katsnelson}},
  \bibinfo{author}{\bibfnamefont{I.~V.} \bibnamefont{Grigorieva}},
  \bibinfo{author}{\bibfnamefont{S.~V.} \bibnamefont{Dubonos}},
  \bibnamefont{and} \bibinfo{author}{\bibfnamefont{A.~A.}
  \bibnamefont{Firsov}}, \bibinfo{journal}{Nature}
  \textbf{\bibinfo{volume}{438}}, \bibinfo{pages}{197} (\bibinfo{year}{2005}).

\bibitem[{\citenamefont{Araki}(2012)}]{araki2012}
\bibinfo{author}{\bibfnamefont{Y.}~\bibnamefont{Araki}},
  \bibinfo{journal}{Physical Review B} \textbf{\bibinfo{volume}{85}},
  \bibinfo{pages}{125436} (\bibinfo{year}{2012}).

\bibitem[{\citenamefont{Katsnelson et~al.}(2006)\citenamefont{Katsnelson,
  Novoselov, and Geim}}]{katsnelson2006_1}
\bibinfo{author}{\bibfnamefont{M.~I.} \bibnamefont{Katsnelson}},
  \bibinfo{author}{\bibfnamefont{K.~S.} \bibnamefont{Novoselov}},
  \bibnamefont{and} \bibinfo{author}{\bibfnamefont{A.~K.} \bibnamefont{Geim}},
  \bibinfo{journal}{Nature Physics} \textbf{\bibinfo{volume}{2}},
  \bibinfo{pages}{620} (\bibinfo{year}{2006}).

\bibitem[{\citenamefont{Jacak et~al.}(2012)\citenamefont{Jacak, Gonczarek,
  Jacak, and J{\' o}{\' z}wiak}}]{jacak2012}
\bibinfo{author}{\bibfnamefont{J.}~\bibnamefont{Jacak}},
  \bibinfo{author}{\bibfnamefont{R.}~\bibnamefont{Gonczarek}},
  \bibinfo{author}{\bibfnamefont{L.}~\bibnamefont{Jacak}}, \bibnamefont{and}
  \bibinfo{author}{\bibfnamefont{I.}~\bibnamefont{J{\' o}{\' z}wiak}},
  \bibinfo{journal}{International Journal of Modern Physics B}
  \textbf{\bibinfo{volume}{29}}, \bibinfo{pages}{1230011}
  (\bibinfo{year}{2012}).

\bibitem[{\citenamefont{Katsnelson}(2006)}]{katsnelson2006_2}
\bibinfo{author}{\bibfnamefont{M.~I.} \bibnamefont{Katsnelson}},
  \bibinfo{journal}{The European Physical Journal B}
  \textbf{\bibinfo{volume}{51}}, \bibinfo{pages}{157} (\bibinfo{year}{2006}).

\bibitem[{\citenamefont{Qu et~al.}(2022)\citenamefont{Qu, Nigge, Link, Levy,
  Michiardi, Spandar, Matth{\'e}, Schneider, Zhdanovich, Starke
  et~al.}}]{qu2022}
\bibinfo{author}{\bibfnamefont{A.~C.} \bibnamefont{Qu}},
  \bibinfo{author}{\bibfnamefont{P.}~\bibnamefont{Nigge}},
  \bibinfo{author}{\bibfnamefont{S.}~\bibnamefont{Link}},
  \bibinfo{author}{\bibfnamefont{G.}~\bibnamefont{Levy}},
  \bibinfo{author}{\bibfnamefont{M.}~\bibnamefont{Michiardi}},
  \bibinfo{author}{\bibfnamefont{P.~L.} \bibnamefont{Spandar}},
  \bibinfo{author}{\bibfnamefont{T.}~\bibnamefont{Matth{\'e}}},
  \bibinfo{author}{\bibfnamefont{M.}~\bibnamefont{Schneider}},
  \bibinfo{author}{\bibfnamefont{S.}~\bibnamefont{Zhdanovich}},
  \bibinfo{author}{\bibfnamefont{U.}~\bibnamefont{Starke}},
  \bibnamefont{et~al.}, \bibinfo{journal}{Science Advances}
  \textbf{\bibinfo{volume}{8}}, \bibinfo{pages}{eabm5180}
  (\bibinfo{year}{2022}).

\bibitem[{\citenamefont{Bao et~al.}(2021)\citenamefont{Bao, Zhang, Zhang, Wu,
  Luo, Zhou, Li, Hou, Yao, Liu et~al.}}]{bao2021}
\bibinfo{author}{\bibfnamefont{C.}~\bibnamefont{Bao}},
  \bibinfo{author}{\bibfnamefont{H.}~\bibnamefont{Zhang}},
  \bibinfo{author}{\bibfnamefont{T.}~\bibnamefont{Zhang}},
  \bibinfo{author}{\bibfnamefont{X.}~\bibnamefont{Wu}},
  \bibinfo{author}{\bibfnamefont{L.}~\bibnamefont{Luo}},
  \bibinfo{author}{\bibfnamefont{S.}~\bibnamefont{Zhou}},
  \bibinfo{author}{\bibfnamefont{Q.}~\bibnamefont{Li}},
  \bibinfo{author}{\bibfnamefont{Y.}~\bibnamefont{Hou}},
  \bibinfo{author}{\bibfnamefont{W.}~\bibnamefont{Yao}},
  \bibinfo{author}{\bibfnamefont{L.}~\bibnamefont{Liu}}, \bibnamefont{et~al.},
  \bibinfo{journal}{Physical Review Letters} \textbf{\bibinfo{volume}{126}},
  \bibinfo{pages}{206804} (\bibinfo{year}{2021}).

\bibitem[{\citenamefont{Guan et~al.}(2024)\citenamefont{Guan, Dutreix, Gonz{\'
  a}lez-Herrero, Ugeda, Brihuega, Katsnelson, Yazyev, and Renard}}]{guan2024}
\bibinfo{author}{\bibfnamefont{Y.}~\bibnamefont{Guan}},
  \bibinfo{author}{\bibfnamefont{C.}~\bibnamefont{Dutreix}},
  \bibinfo{author}{\bibfnamefont{H.}~\bibnamefont{Gonz{\' a}lez-Herrero}},
  \bibinfo{author}{\bibfnamefont{M.~M.} \bibnamefont{Ugeda}},
  \bibinfo{author}{\bibfnamefont{I.}~\bibnamefont{Brihuega}},
  \bibinfo{author}{\bibfnamefont{M.~I.} \bibnamefont{Katsnelson}},
  \bibinfo{author}{\bibfnamefont{O.~V.} \bibnamefont{Yazyev}},
  \bibnamefont{and} \bibinfo{author}{\bibfnamefont{V.~T.}
  \bibnamefont{Renard}}, \bibinfo{journal}{Nature Communications}
  \textbf{\bibinfo{volume}{15}}, \bibinfo{pages}{2927} (\bibinfo{year}{2024}).

\bibitem[{\citenamefont{Lin et~al.}(2017)\citenamefont{Lin, Qin, Zeng, Chen,
  Cui, Cho, Qiao, and Zhang}}]{lin2017}
\bibinfo{author}{\bibfnamefont{Z.}~\bibnamefont{Lin}},
  \bibinfo{author}{\bibfnamefont{W.}~\bibnamefont{Qin}},
  \bibinfo{author}{\bibfnamefont{J.}~\bibnamefont{Zeng}},
  \bibinfo{author}{\bibfnamefont{W.}~\bibnamefont{Chen}},
  \bibinfo{author}{\bibfnamefont{P.}~\bibnamefont{Cui}},
  \bibinfo{author}{\bibfnamefont{J.~H.} \bibnamefont{Cho}},
  \bibinfo{author}{\bibfnamefont{Z.}~\bibnamefont{Qiao}}, \bibnamefont{and}
  \bibinfo{author}{\bibfnamefont{Z.}~\bibnamefont{Zhang}},
  \bibinfo{journal}{Nano Letters} \textbf{\bibinfo{volume}{17}},
  \bibinfo{pages}{4013} (\bibinfo{year}{2017}).

\bibitem[{\citenamefont{Wu and Hu}(2016)}]{wu2016}
\bibinfo{author}{\bibfnamefont{L.~H.} \bibnamefont{Wu}} \bibnamefont{and}
  \bibinfo{author}{\bibfnamefont{X.}~\bibnamefont{Hu}},
  \bibinfo{journal}{Scientific Reports} \textbf{\bibinfo{volume}{6}},
  \bibinfo{pages}{24347} (\bibinfo{year}{2016}).

\bibitem[{\citenamefont{Liu et~al.}(2017)\citenamefont{Liu, Lian, Li, Xu, and
  Duan}}]{liu2017}
\bibinfo{author}{\bibfnamefont{Y.}~\bibnamefont{Liu}},
  \bibinfo{author}{\bibfnamefont{C.~S.} \bibnamefont{Lian}},
  \bibinfo{author}{\bibfnamefont{Y.}~\bibnamefont{Li}},
  \bibinfo{author}{\bibfnamefont{Y.}~\bibnamefont{Xu}}, \bibnamefont{and}
  \bibinfo{author}{\bibfnamefont{W.}~\bibnamefont{Duan}},
  \bibinfo{journal}{Physical Review Letters} \textbf{\bibinfo{volume}{119}},
  \bibinfo{pages}{255901} (\bibinfo{year}{2017}).

\bibitem[{\citenamefont{Garc{\'i}a et~al.}(2024)\citenamefont{Garc{\'i}a,
  Betancur-Ocampo, S{\'a}nchez-Ochoa, and Stegmann}}]{garcia2024}
\bibinfo{author}{\bibfnamefont{S.~G.} \bibnamefont{Garc{\'i}a}},
  \bibinfo{author}{\bibfnamefont{Y.}~\bibnamefont{Betancur-Ocampo}},
  \bibinfo{author}{\bibfnamefont{F.}~\bibnamefont{S{\'a}nchez-Ochoa}},
  \bibnamefont{and} \bibinfo{author}{\bibfnamefont{T.}~\bibnamefont{Stegmann}},
  \bibinfo{journal}{Nano Letters} \textbf{\bibinfo{volume}{24}},
  \bibinfo{pages}{2322} (\bibinfo{year}{2024}).

\bibitem[{\citenamefont{Szcz\c{e}\'{s}niak and
  Szcz\c{e}\'{s}niak}(2019)}]{szczesniak2019}
\bibinfo{author}{\bibfnamefont{D.}~\bibnamefont{Szcz\c{e}\'{s}niak}}
  \bibnamefont{and}
  \bibinfo{author}{\bibfnamefont{R.}~\bibnamefont{Szcz\c{e}\'{s}niak}},
  \bibinfo{journal}{Physical Review B} \textbf{\bibinfo{volume}{99}},
  \bibinfo{pages}{224512} (\bibinfo{year}{2019}).

\bibitem[{\citenamefont{Profeta et~al.}(2012)\citenamefont{Profeta, Calandra,
  and Mauri}}]{profeta2012}
\bibinfo{author}{\bibfnamefont{G.}~\bibnamefont{Profeta}},
  \bibinfo{author}{\bibfnamefont{M.}~\bibnamefont{Calandra}}, \bibnamefont{and}
  \bibinfo{author}{\bibfnamefont{F.}~\bibnamefont{Mauri}},
  \bibinfo{journal}{Nature Physics} \textbf{\bibinfo{volume}{8}},
  \bibinfo{pages}{131} (\bibinfo{year}{2012}).

\bibitem[{\citenamefont{Ludbrook et~al.}(2015)\citenamefont{Ludbrook, Levy,
  Nigge, Zonno, Schneider, Dvorak, Veenstra, Zhdanovich, Wong, Dosanjh
  et~al.}}]{ludbrook2015}
\bibinfo{author}{\bibfnamefont{B.~M.} \bibnamefont{Ludbrook}},
  \bibinfo{author}{\bibfnamefont{G.}~\bibnamefont{Levy}},
  \bibinfo{author}{\bibfnamefont{P.}~\bibnamefont{Nigge}},
  \bibinfo{author}{\bibfnamefont{M.}~\bibnamefont{Zonno}},
  \bibinfo{author}{\bibfnamefont{M.}~\bibnamefont{Schneider}},
  \bibinfo{author}{\bibfnamefont{D.~J.} \bibnamefont{Dvorak}},
  \bibinfo{author}{\bibfnamefont{C.~N.} \bibnamefont{Veenstra}},
  \bibinfo{author}{\bibfnamefont{S.}~\bibnamefont{Zhdanovich}},
  \bibinfo{author}{\bibfnamefont{D.}~\bibnamefont{Wong}},
  \bibinfo{author}{\bibfnamefont{P.}~\bibnamefont{Dosanjh}},
  \bibnamefont{et~al.}, \bibinfo{journal}{PNAS} \textbf{\bibinfo{volume}{112}},
  \bibinfo{pages}{11795} (\bibinfo{year}{2015}).

\bibitem[{\citenamefont{Eom and Koo}(2020)}]{eom2020}
\bibinfo{author}{\bibfnamefont{D.}~\bibnamefont{Eom}} \bibnamefont{and}
  \bibinfo{author}{\bibfnamefont{J.~Y.} \bibnamefont{Koo}},
  \bibinfo{journal}{Nanoscale} \textbf{\bibinfo{volume}{12}},
  \bibinfo{pages}{19604} (\bibinfo{year}{2020}).

\bibitem[{\citenamefont{Sorella et~al.}(2018)\citenamefont{Sorella, Seki,
  Brovko, Shirakawa, Miyakoshi, Yunoki, and Tosatti}}]{sorella2018}
\bibinfo{author}{\bibfnamefont{S.}~\bibnamefont{Sorella}},
  \bibinfo{author}{\bibfnamefont{K.}~\bibnamefont{Seki}},
  \bibinfo{author}{\bibfnamefont{O.~O.} \bibnamefont{Brovko}},
  \bibinfo{author}{\bibfnamefont{T.}~\bibnamefont{Shirakawa}},
  \bibinfo{author}{\bibfnamefont{S.}~\bibnamefont{Miyakoshi}},
  \bibinfo{author}{\bibfnamefont{S.}~\bibnamefont{Yunoki}}, \bibnamefont{and}
  \bibinfo{author}{\bibfnamefont{E.}~\bibnamefont{Tosatti}},
  \bibinfo{journal}{Physical Review Letters} \textbf{\bibinfo{volume}{121}},
  \bibinfo{pages}{066402} (\bibinfo{year}{2018}).

\bibitem[{\citenamefont{Zhang et~al.}(2022)\citenamefont{Zhang, Bao, Sch{\"
  u}ler, Zhou, Li, Luo, Yao, Wang, Devereaux, and Zhou}}]{zhang2021}
\bibinfo{author}{\bibfnamefont{H.}~\bibnamefont{Zhang}},
  \bibinfo{author}{\bibfnamefont{C.}~\bibnamefont{Bao}},
  \bibinfo{author}{\bibfnamefont{M.}~\bibnamefont{Sch{\" u}ler}},
  \bibinfo{author}{\bibfnamefont{S.}~\bibnamefont{Zhou}},
  \bibinfo{author}{\bibfnamefont{Q.}~\bibnamefont{Li}},
  \bibinfo{author}{\bibfnamefont{L.}~\bibnamefont{Luo}},
  \bibinfo{author}{\bibfnamefont{W.}~\bibnamefont{Yao}},
  \bibinfo{author}{\bibfnamefont{Z.}~\bibnamefont{Wang}},
  \bibinfo{author}{\bibfnamefont{T.~P.} \bibnamefont{Devereaux}},
  \bibnamefont{and} \bibinfo{author}{\bibfnamefont{S.}~\bibnamefont{Zhou}},
  \bibinfo{journal}{National Science Review} \textbf{\bibinfo{volume}{9}},
  \bibinfo{pages}{nwab175} (\bibinfo{year}{2022}).

\bibitem[{\citenamefont{Costa et~al.}(2024)\citenamefont{Costa, Cohen-Stead,
  and Johnston}}]{costa2024}
\bibinfo{author}{\bibfnamefont{S.~M.} \bibnamefont{Costa}},
  \bibinfo{author}{\bibfnamefont{B.}~\bibnamefont{Cohen-Stead}},
  \bibnamefont{and} \bibinfo{author}{\bibfnamefont{S.}~\bibnamefont{Johnston}},
  \bibinfo{journal}{Physical Review B} \textbf{\bibinfo{volume}{110}},
  \bibinfo{pages}{115130} (\bibinfo{year}{2024}).

\bibitem[{\citenamefont{Otsuka and Yunoki}(2024)}]{otsuka2024}
\bibinfo{author}{\bibfnamefont{Y.}~\bibnamefont{Otsuka}} \bibnamefont{and}
  \bibinfo{author}{\bibfnamefont{S.}~\bibnamefont{Yunoki}},
  \bibinfo{journal}{Physical Review B} \textbf{\bibinfo{volume}{109}},
  \bibinfo{pages}{115131} (\bibinfo{year}{2024}).

\bibitem[{\citenamefont{Krok et~al.}(2023)\citenamefont{Krok, Adamczyk,
  Durajski, and Szcz\c{e}\'{s}niak}}]{krok2023}
\bibinfo{author}{\bibfnamefont{K.~A.} \bibnamefont{Krok}},
  \bibinfo{author}{\bibfnamefont{M.~M.} \bibnamefont{Adamczyk}},
  \bibinfo{author}{\bibfnamefont{A.~P.} \bibnamefont{Durajski}},
  \bibnamefont{and}
  \bibinfo{author}{\bibfnamefont{R.}~\bibnamefont{Szcz\c{e}\'{s}niak}},
  \bibinfo{journal}{Physical Review B} \textbf{\bibinfo{volume}{108}},
  \bibinfo{pages}{054512} (\bibinfo{year}{2023}).

\bibitem[{\citenamefont{{Pe{\v{s}}i{\'{c}}}
  et~al.}(2014)\citenamefont{{Pe{\v{s}}i{\'{c}}}, Gaji{\' c}, Hingerl, and
  Beli{\' c}}}]{pesic2014}
\bibinfo{author}{\bibfnamefont{J.}~\bibnamefont{{Pe{\v{s}}i{\'{c}}}}},
  \bibinfo{author}{\bibfnamefont{R.}~\bibnamefont{Gaji{\' c}}},
  \bibinfo{author}{\bibfnamefont{K.}~\bibnamefont{Hingerl}}, \bibnamefont{and}
  \bibinfo{author}{\bibfnamefont{M.}~\bibnamefont{Beli{\' c}}},
  \bibinfo{journal}{EPL} \textbf{\bibinfo{volume}{108}}, \bibinfo{pages}{67005}
  (\bibinfo{year}{2014}).

\bibitem[{\citenamefont{Kaloni et~al.}(2013)\citenamefont{Kaloni, Balatsky, and
  Schwingenschl{\"o}gl}}]{kaloni2013}
\bibinfo{author}{\bibfnamefont{T.~P.} \bibnamefont{Kaloni}},
  \bibinfo{author}{\bibfnamefont{A.~V.} \bibnamefont{Balatsky}},
  \bibnamefont{and}
  \bibinfo{author}{\bibfnamefont{U.}~\bibnamefont{Schwingenschl{\"o}gl}},
  \bibinfo{journal}{EPL} \textbf{\bibinfo{volume}{104}}, \bibinfo{pages}{47013}
  (\bibinfo{year}{2013}).

\bibitem[{\citenamefont{Szcz{\c e}{\' s}niak}(2023)}]{szczesniak2023}
\bibinfo{author}{\bibfnamefont{D.}~\bibnamefont{Szcz{\c e}{\' s}niak}},
  \bibinfo{journal}{EPL} \textbf{\bibinfo{volume}{142}}, \bibinfo{pages}{36002}
  (\bibinfo{year}{2023}).

\bibitem[{\citenamefont{{Castro Neto} and Guinea}(2007)}]{castro2007}
\bibinfo{author}{\bibfnamefont{A.~H.} \bibnamefont{{Castro Neto}}}
  \bibnamefont{and} \bibinfo{author}{\bibfnamefont{F.}~\bibnamefont{Guinea}},
  \bibinfo{journal}{Physical Review B} \textbf{\bibinfo{volume}{75}},
  \bibinfo{pages}{045404} (\bibinfo{year}{2007}).

\bibitem[{\citenamefont{Cappelluti and Profeta}(2012)}]{cappelluti2012}
\bibinfo{author}{\bibfnamefont{E.}~\bibnamefont{Cappelluti}} \bibnamefont{and}
  \bibinfo{author}{\bibfnamefont{G.}~\bibnamefont{Profeta}},
  \bibinfo{journal}{Physical Review B} \textbf{\bibinfo{volume}{85}},
  \bibinfo{pages}{205436} (\bibinfo{year}{2012}).

\bibitem[{\citenamefont{Khan and Allen}(1984)}]{khan1984}
\bibinfo{author}{\bibfnamefont{F.~S.} \bibnamefont{Khan}} \bibnamefont{and}
  \bibinfo{author}{\bibfnamefont{P.~B.} \bibnamefont{Allen}},
  \bibinfo{journal}{Physical Review B} \textbf{\bibinfo{volume}{29}},
  \bibinfo{pages}{3341} (\bibinfo{year}{1984}).

\bibitem[{\citenamefont{Gr{\" u}neis et~al.}(2008)\citenamefont{Gr{\" u}neis,
  Attaccalite, Wirtz, Shiozawa, Saito, Pichler, and Rubio}}]{gruneis2008}
\bibinfo{author}{\bibfnamefont{A.}~\bibnamefont{Gr{\" u}neis}},
  \bibinfo{author}{\bibfnamefont{C.}~\bibnamefont{Attaccalite}},
  \bibinfo{author}{\bibfnamefont{L.}~\bibnamefont{Wirtz}},
  \bibinfo{author}{\bibfnamefont{H.}~\bibnamefont{Shiozawa}},
  \bibinfo{author}{\bibfnamefont{R.}~\bibnamefont{Saito}},
  \bibinfo{author}{\bibfnamefont{T.}~\bibnamefont{Pichler}}, \bibnamefont{and}
  \bibinfo{author}{\bibfnamefont{A.}~\bibnamefont{Rubio}},
  \bibinfo{journal}{Physical Review B} \textbf{\bibinfo{volume}{78}},
  \bibinfo{pages}{205425} (\bibinfo{year}{2008}).

\bibitem[{\citenamefont{Pereira et~al.}(2009)\citenamefont{Pereira, Neto, and
  Peres}}]{pereira2009}
\bibinfo{author}{\bibfnamefont{V.~M.} \bibnamefont{Pereira}},
  \bibinfo{author}{\bibfnamefont{A.~H.~C.} \bibnamefont{Neto}},
  \bibnamefont{and} \bibinfo{author}{\bibfnamefont{N.~M.~R.}
  \bibnamefont{Peres}}, \bibinfo{journal}{Physical Review B}
  \textbf{\bibinfo{volume}{80}}, \bibinfo{pages}{045401}
  (\bibinfo{year}{2009}).

\bibitem[{\citenamefont{Naumis et~al.}(2017)\citenamefont{Naumis,
  Barraza-Lopez, Oliva-Leyva, and Terrones}}]{naumis2017}
\bibinfo{author}{\bibfnamefont{G.~G.} \bibnamefont{Naumis}},
  \bibinfo{author}{\bibfnamefont{S.}~\bibnamefont{Barraza-Lopez}},
  \bibinfo{author}{\bibfnamefont{M.}~\bibnamefont{Oliva-Leyva}},
  \bibnamefont{and} \bibinfo{author}{\bibfnamefont{H.}~\bibnamefont{Terrones}},
  \bibinfo{journal}{Reports on Progress in Physics}
  \textbf{\bibinfo{volume}{80}}, \bibinfo{pages}{096501}
  (\bibinfo{year}{2017}).

\bibitem[{\citenamefont{Levenberg}(1944)}]{levenberg1944}
\bibinfo{author}{\bibfnamefont{K.}~\bibnamefont{Levenberg}},
  \bibinfo{journal}{Quarterly of Applied Mathematics}
  \textbf{\bibinfo{volume}{2}}, \bibinfo{pages}{164} (\bibinfo{year}{1944}).

\bibitem[{\citenamefont{Marquardt}(1963)}]{marquardt1963}
\bibinfo{author}{\bibfnamefont{D.}~\bibnamefont{Marquardt}},
  \bibinfo{journal}{SIAM Journal on Applied Mathematics}
  \textbf{\bibinfo{volume}{11}}, \bibinfo{pages}{431} (\bibinfo{year}{1963}).

\bibitem[{\citenamefont{Ribeiro et~al.}(2009)\citenamefont{Ribeiro, Pereira,
  Peres, Briddon, and Neto}}]{ribeiro2009}
\bibinfo{author}{\bibfnamefont{R.~M.} \bibnamefont{Ribeiro}},
  \bibinfo{author}{\bibfnamefont{V.~M.} \bibnamefont{Pereira}},
  \bibinfo{author}{\bibfnamefont{N.~M.~R.} \bibnamefont{Peres}},
  \bibinfo{author}{\bibfnamefont{P.~R.} \bibnamefont{Briddon}},
  \bibnamefont{and} \bibinfo{author}{\bibfnamefont{A.~H.~C.}
  \bibnamefont{Neto}}, \bibinfo{journal}{New Journal of Physics}
  \textbf{\bibinfo{volume}{11}}, \bibinfo{pages}{115002}
  (\bibinfo{year}{2009}).

\bibitem[{\citenamefont{Kundu}(2011)}]{kundu2009}
\bibinfo{author}{\bibfnamefont{R.}~\bibnamefont{Kundu}},
  \bibinfo{journal}{Modern Physics Letters B} \textbf{\bibinfo{volume}{25}},
  \bibinfo{pages}{163} (\bibinfo{year}{2011}).

\bibitem[{\citenamefont{Botello-M{\'e}ndez
  et~al.}(2018)\citenamefont{Botello-M{\'e}ndez, Obeso-Jureidini, and
  Naumis}}]{botello-mendez2018}
\bibinfo{author}{\bibfnamefont{A.~R.} \bibnamefont{Botello-M{\'e}ndez}},
  \bibinfo{author}{\bibfnamefont{J.~C.} \bibnamefont{Obeso-Jureidini}},
  \bibnamefont{and} \bibinfo{author}{\bibfnamefont{G.~G.}
  \bibnamefont{Naumis}}, \bibinfo{journal}{The Journal of Physical Chemistry C}
  \textbf{\bibinfo{volume}{122}}, \bibinfo{pages}{15753}
  (\bibinfo{year}{2018}).

\bibitem[{\citenamefont{Reich et~al.}(2002)\citenamefont{Reich, Maultzsch,
  Thomsen, and Ordej{\'o}n}}]{reich2002}
\bibinfo{author}{\bibfnamefont{S.}~\bibnamefont{Reich}},
  \bibinfo{author}{\bibfnamefont{J.}~\bibnamefont{Maultzsch}},
  \bibinfo{author}{\bibfnamefont{C.}~\bibnamefont{Thomsen}}, \bibnamefont{and}
  \bibinfo{author}{\bibfnamefont{P.}~\bibnamefont{Ordej{\'o}n}},
  \bibinfo{journal}{Physical Review B} \textbf{\bibinfo{volume}{66}},
  \bibinfo{pages}{035412} (\bibinfo{year}{2002}).

\bibitem[{\citenamefont{Farjam and Rafii-Tabar}(2009)}]{farjam2009}
\bibinfo{author}{\bibfnamefont{M.}~\bibnamefont{Farjam}} \bibnamefont{and}
  \bibinfo{author}{\bibfnamefont{H.}~\bibnamefont{Rafii-Tabar}},
  \bibinfo{journal}{Physical Review B} \textbf{\bibinfo{volume}{79}},
  \bibinfo{pages}{045417} (\bibinfo{year}{2009}).

\bibitem[{\citenamefont{Slater and Koster}(1954)}]{slater1954}
\bibinfo{author}{\bibfnamefont{J.~C.} \bibnamefont{Slater}} \bibnamefont{and}
  \bibinfo{author}{\bibfnamefont{G.~F.} \bibnamefont{Koster}},
  \bibinfo{journal}{Physical Review} \textbf{\bibinfo{volume}{94}},
  \bibinfo{pages}{1498} (\bibinfo{year}{1954}).

\bibitem[{\citenamefont{Samsonidze et~al.}(2007)\citenamefont{Samsonidze,
  Barros, Saito, Jiang, Dresselhaus, and Dresselhaus}}]{samsonidze2007}
\bibinfo{author}{\bibfnamefont{G.~G.} \bibnamefont{Samsonidze}},
  \bibinfo{author}{\bibfnamefont{E.~B.} \bibnamefont{Barros}},
  \bibinfo{author}{\bibfnamefont{R.}~\bibnamefont{Saito}},
  \bibinfo{author}{\bibfnamefont{J.}~\bibnamefont{Jiang}},
  \bibinfo{author}{\bibfnamefont{G.}~\bibnamefont{Dresselhaus}},
  \bibnamefont{and} \bibinfo{author}{\bibfnamefont{M.~S.}
  \bibnamefont{Dresselhaus}}, \bibinfo{journal}{Physical Review B}
  \textbf{\bibinfo{volume}{75}}, \bibinfo{pages}{155420}
  (\bibinfo{year}{2007}).

\bibitem[{\citenamefont{Lazzeri et~al.}(2008)\citenamefont{Lazzeri,
  Attaccalite, Wirtz, and Mauri}}]{lazzeri2008}
\bibinfo{author}{\bibfnamefont{M.}~\bibnamefont{Lazzeri}},
  \bibinfo{author}{\bibfnamefont{C.}~\bibnamefont{Attaccalite}},
  \bibinfo{author}{\bibfnamefont{L.}~\bibnamefont{Wirtz}}, \bibnamefont{and}
  \bibinfo{author}{\bibfnamefont{F.}~\bibnamefont{Mauri}},
  \bibinfo{journal}{Physical Review B} \textbf{\bibinfo{volume}{78}},
  \bibinfo{pages}{081406(R)} (\bibinfo{year}{2008}).

\bibitem[{\citenamefont{Wan et~al.}(2013)\citenamefont{Wan, Ge, Yang, and
  Yao}}]{wan2013}
\bibinfo{author}{\bibfnamefont{W.}~\bibnamefont{Wan}},
  \bibinfo{author}{\bibfnamefont{Y.}~\bibnamefont{Ge}},
  \bibinfo{author}{\bibfnamefont{F.}~\bibnamefont{Yang}}, \bibnamefont{and}
  \bibinfo{author}{\bibfnamefont{Y.}~\bibnamefont{Yao}}, \bibinfo{journal}{EPL}
  \textbf{\bibinfo{volume}{104}}, \bibinfo{pages}{36001}
  (\bibinfo{year}{2013}).

\bibitem[{\citenamefont{Zhou et~al.}(2015)\citenamefont{Zhou, Sun, Wang, and
  Jena}}]{zhou2015}
\bibinfo{author}{\bibfnamefont{J.}~\bibnamefont{Zhou}},
  \bibinfo{author}{\bibfnamefont{Q.}~\bibnamefont{Sun}},
  \bibinfo{author}{\bibfnamefont{Q.}~\bibnamefont{Wang}}, \bibnamefont{and}
  \bibinfo{author}{\bibfnamefont{P.}~\bibnamefont{Jena}},
  \bibinfo{journal}{Physical Review B} \textbf{\bibinfo{volume}{92}},
  \bibinfo{pages}{064505} (\bibinfo{year}{2015}).

\bibitem[{\citenamefont{Khoa et~al.}(2019)\citenamefont{Khoa, Nguyen, Bui,
  Phuc, Hoi, Hieu, Nha, Huynh, Nhan, and Hieu}}]{khoa2019}
\bibinfo{author}{\bibfnamefont{D.~Q.} \bibnamefont{Khoa}},
  \bibinfo{author}{\bibfnamefont{C.~V.} \bibnamefont{Nguyen}},
  \bibinfo{author}{\bibfnamefont{L.~M.} \bibnamefont{Bui}},
  \bibinfo{author}{\bibfnamefont{H.~V.} \bibnamefont{Phuc}},
  \bibinfo{author}{\bibfnamefont{B.~D.} \bibnamefont{Hoi}},
  \bibinfo{author}{\bibfnamefont{N.~V.} \bibnamefont{Hieu}},
  \bibinfo{author}{\bibfnamefont{V.~Q.} \bibnamefont{Nha}},
  \bibinfo{author}{\bibfnamefont{N.}~\bibnamefont{Huynh}},
  \bibinfo{author}{\bibfnamefont{L.~C.} \bibnamefont{Nhan}}, \bibnamefont{and}
  \bibinfo{author}{\bibfnamefont{N.~N.} \bibnamefont{Hieu}},
  \bibinfo{journal}{Materials Research Express} \textbf{\bibinfo{volume}{6}},
  \bibinfo{pages}{045605} (\bibinfo{year}{2019}).

\bibitem[{\citenamefont{Park and Choi}(2015)}]{park2015}
\bibinfo{author}{\bibfnamefont{J.~S.} \bibnamefont{Park}} \bibnamefont{and}
  \bibinfo{author}{\bibfnamefont{H.~J.} \bibnamefont{Choi}},
  \bibinfo{journal}{Physical Review B} \textbf{\bibinfo{volume}{92}},
  \bibinfo{pages}{045402} (\bibinfo{year}{2015}).

\end{thebibliography}

\end{document}